\documentclass[12pt,a4paper]{article}
\usepackage{amsmath,amssymb,url}
\usepackage{graphicx,tabularx}
\usepackage{array,dsfont}
\usepackage{cleveref}
\usepackage{color}

\crefformat{equation}{Eq.~(#2#1#3)}
\crefformat{section}{Section~#2#1#3}
\crefformat{figure}{Figure~#2#1#3}
\crefformat{table}{Table~#2#1#3}
\crefformat{chapter}{Chapter~#2#1#3}
\crefformat{appendix}{Appendix~#2#1#3}

\makeatletter
\newif\if@preliminary
\@preliminaryfalse
\def\preliminary{\@preliminarytrue}
\def\preprintno#1{\def\@preprintno{#1}}
\def\address#1{\def\@address{#1}}
\def\email#1#2{\thanks{\tt #1@{}#2}}
\def\abstract#1{\def\@abstract{#1}}
\renewcommand\abstractname{ABSTRACT}
\newlength\preprintnoskip
\newlength\abstractwidth
\@titlepagetrue
\renewcommand\maketitle{\begin{titlepage}%
  \let\footnotesize\small
  \hfill\parbox{\preprintnoskip}{%
  \begin{flushright}\@preprintno\end{flushright}}\hspace*{1cm}
  \vskip 60\p@
  \begin{center}%
    {\Large\bf\boldmath \@title \par}\vskip 1cm%
    {\sc\@author \par}\vskip 3mm%
    {\@address \par}%
    \if@preliminary
      \vskip 2cm {\large\sf PRELIMINARY DRAFT \par \@date}%
    \fi
  \end{center}\par
  \@thanks
  \vfill
  \begin{center}%
    \parbox{\abstractwidth}{\centerline{\abstractname}%
    \vskip 3mm%
    \@abstract}
  \end{center}
  \end{titlepage}%
  \setcounter{footnote}{0}%
  \let\thanks\relax\let\maketitle\relax
  \gdef\@thanks{}\gdef\@author{}\gdef\@address{}%
  \gdef\@title{}\gdef\@abstract{}\gdef\@preprintno{}
}%
\def\@citex[#1]#2{\if@filesw\immediate\write\@auxout{\string\citation{#2}}\fi
  \def\@citea{}\@cite{\@for\@citeb:=#2\do
    {\@citea\def\@citea{,\penalty\@m}\@ifundefined
       {b@\@citeb}{{\bf ?}\@warning
       {Citation `\@citeb' on page \thepage \space undefined}}%
\hbox{\csname b@\@citeb\endcsname}}}{#1}}
\def\citerange{\@ifnextchar [{\@tempswatrue\@citexr}{\@tempswafalse\@citexr[]}}
\def\@citexr[#1]#2{\if@filesw\immediate\write\@auxout{\string\citation{#2}}\fi
  \def\@citea{}\@cite{\@for\@citeb:=#2\do
    {\@citea\def\@citea{--\penalty\@m}\@ifundefined
       {b@\@citeb}{{\bf ?}\@warning
       {Citation `\@citeb' on page \thepage \space undefined}}%
\hbox{\csname b@\@citeb\endcsname}}}{#1}}
\long\def\@makecaption#1#2{%
  \sbox\@tempboxa{#1: \emph{#2}}%
  \ifdim \wd\@tempboxa >\hsize
    #1: \emph{#2}\par
  \else
    \hbox to\hsize{\hfil\box\@tempboxa\hfil}%
  \fi
  \vskip\belowcaptionskip}
\def\fmslash{\@ifnextchar[{\fmsl@sh}{\fmsl@sh[0mu]}}
\def\fmsl@sh[#1]#2{%
  \mathchoice
    {\@fmsl@sh\displaystyle{#1}{#2}}%
    {\@fmsl@sh\textstyle{#1}{#2}}%
    {\@fmsl@sh\scriptstyle{#1}{#2}}%
    {\@fmsl@sh\scriptscriptstyle{#1}{#2}}}
\def\@fmsl@sh#1#2#3{\m@th\ooalign{$\hfil#1\mkern#2/\hfil$\crcr$#1#3$}}
\makeatother

\newcommand\ltap{\
  \raise.3ex\hbox{$<$\kern-.75em\lower1ex\hbox{$\sim$}}\ }
\newcommand\gtap{\
  \raise.3ex\hbox{$>$\kern-.75em\lower1ex\hbox{$\sim$}}\ }

\newcommand\simge{\mathrel{%
   \rlap{\raise 0.511ex \hbox{$>$}}{\lower 0.511ex \hbox{$\sim$}}}}
\newcommand\simle{\mathrel{
   \rlap{\raise 0.511ex \hbox{$<$}}{\lower 0.511ex \hbox{$\sim$}}}}

\newcommand\be{\begin{equation}}
\newcommand\ee{\end{equation}}
\newcommand\bea{\begin{eqnarray}}
\newcommand\eea{\end{eqnarray}}
\newcommand\ba{\begin{array}}
\newcommand\ea{\end{array}}

\def\bq{\begin{equation}}
\def\eq{\end{equation}}
\def\ba{\begin{eqnarray}}
\def\ea{\end{eqnarray}}

\begin{document}

\date{\today}

\preprintno{DESY 16-031}

\title{The BSM Physics Case of the ILC}

\author{J\"urgen Reuter\email{juergen.reuter}{desy.de}$^{a,\ast}$
}

\address{\it%
$^a$DESY Theory Group, \\
  Notkestr. 85, D-22607 Hamburg, Germany
\\[3\baselineskip]
$^\ast$ Talk presented at the International Workshop on Future Linear
Colliders (LCWS15), Whistler, Canada, 2-6 November 2015
}

\abstract{
In this talk I summarize the physics case of the International Linear
Collider (ILC) focusing on its potential towards discovery,
discrimation or disentanglement of new physics beyond the Standard
Model (BSM).}

\maketitle


\section{The physics case of the International Linear Collider}

For this talk, I was asked to collect material for the physics case of
the planned International Linear Collider (ILC) regarding its
discovery potential for New Physics beyond the Standard Model. The
physics case of a lepton collider with energies up to $\sqrt{s} = 500$
GeV (and beyond) has been outlined many times, starting from workshops
during summer 1987 at KEK~\cite{Hagiwara:1987df}, La
Thuile~\cite{Tracas:1987ud} and SLAC~\cite{Peskin:1988cz}, following
the TESLA Technical Design
Report~\cite{Richard:2001qm,AguilarSaavedra:2001rg}, towards the ILC
Technical Design Report/Detailed Baseline
Design~\cite{Behnke:2013xla,Baer:2013cma}. Shortly after that, the
physics case has been updated for the latest U.S. Snowmass community
summer study~\cite{Baer:2013vqa}. In 2015, the ILC Physics Working
Group was asked to provide a condensed version of the physics
case~\cite{Fujii:2015jha} for the expert committees advising the
Japanese Ministry for Education, Science, and Technology, MEXT.
Though being a member of this Physics Working Group, this document is
no official statement by them. The Physics case
document~\cite{Fujii:2015jha} is accompanied by more definite running
scenarios for the machine, given both technical machine and physics
aspects, which have been worked out~\cite{Brau:2015ppa}, and which
allow to assess the development in precision of measurements of all
parameters with running time of the machine. 

\begin{figure}
  \begin{center}
    \includegraphics[width=.58\textwidth]{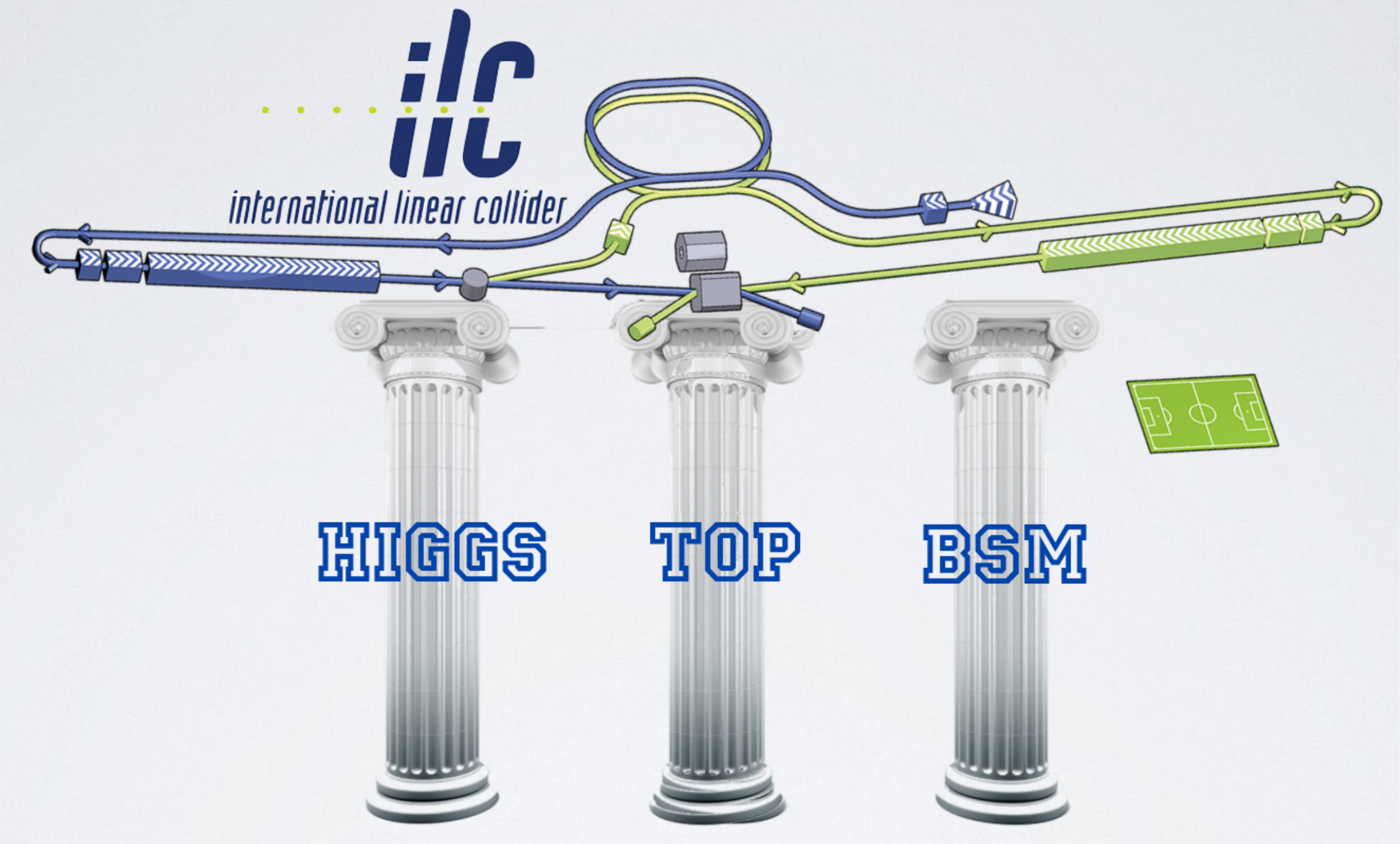}
    \includegraphics[width=.4\textwidth]{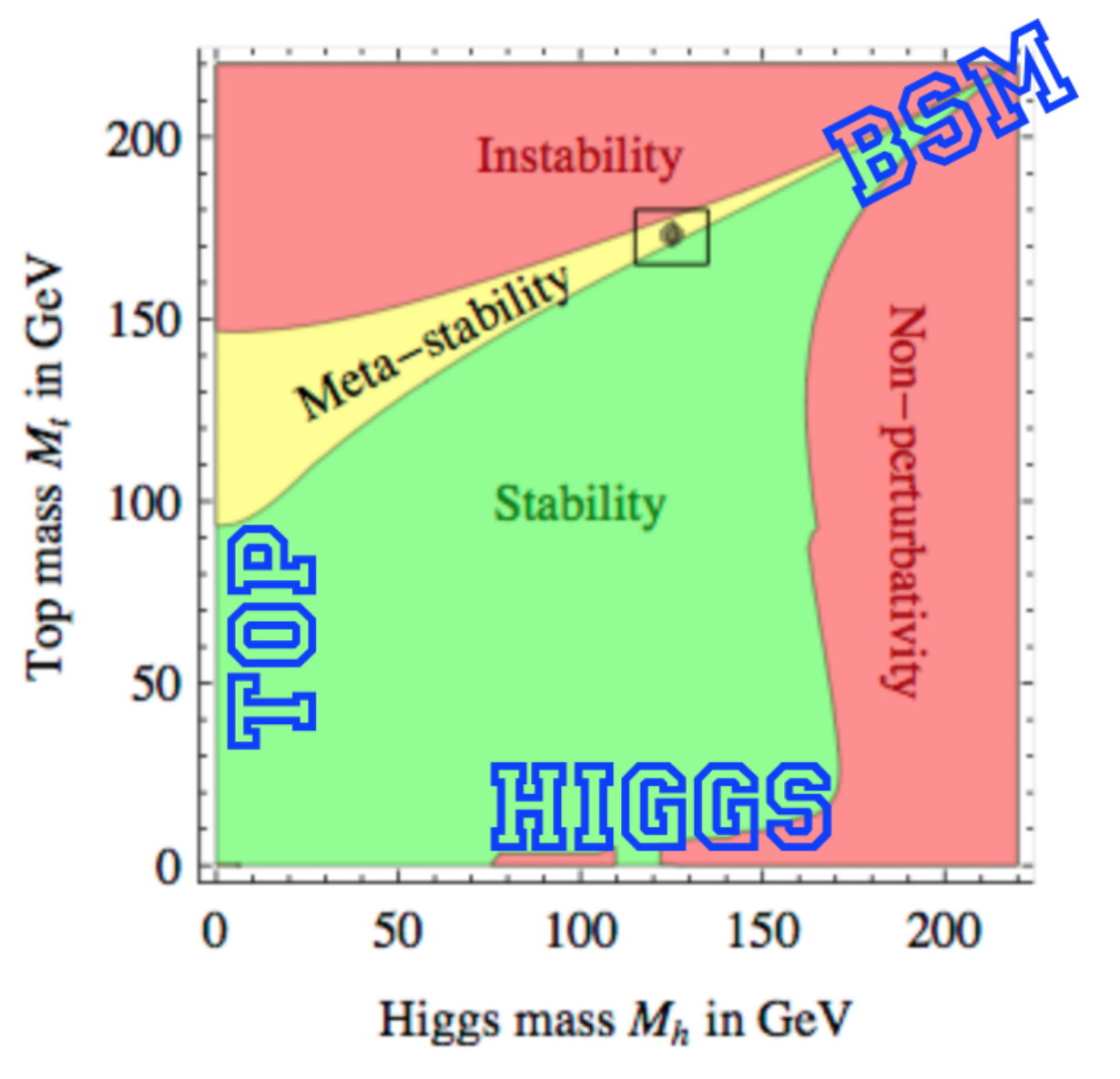}    
  \end{center}
  \caption{\label{fig:pillars} Left: The physics programme of a
    high-energy $e^+e^-$ machine like the ILC rests on three pillars:
    its Higgs physics and top physics programmes as well as its search
    potential for new physics (BSM). Right: stability diagram of the
    electroweak vacuum, depending on top, Higgs, and BSM physics, base
    plot from~\cite{Degrassi:2012ry}.}  
\end{figure}

The physics programme of any future high-energy lepton collider that
reaches at least 500 GeV rests on three pillars,
cf. the left-hand side of Fig.~\ref{fig:pillars}: its precision Higgs
programme, its precision top quark programme and the direct search
potential for BSM physics. These are motivated most effectively by the
question of the microscopic structure behind the electroweak sector of
the SM and its vacuum structure (right hand side of
Fig.~\ref{fig:pillars}). To determine the stability properties of the
electroweak vacuum, precision measurements of the main ingredients,
the Higgs boson and the top quark are necessary, and particularly
their interplay, namely the top Yukawa coupling. Since this plot has
been made under the assumption that there is no BSM up to the Planck
scale (at least not with electroweak quantum numbers), this question
also depends on the BSM paradigm.

In the field of particle physics, over the years, a paradigm has
developed, which I would consider clearly wrong, namely that hadron
machines are discovery machines, while lepton colliders are precision
machines. First of all, looking particularly at the vast number of
impressive measurements on top, $W$, $Z$, and Higgs physics from the
LHC experiments ATLAS and CMS, it is clear, that also hadron colliders
can serve as precision machines. On the other hand, most of the
spectacular discoveries, especially the unexpected ones, have been
made with lepton beams: starting from the revelation of quark
substructure of hadrons in DIS experiements at SLAC in 1969 with
electron beams, over the neutral current discovery by the Gargamelle
experiment at CERN 1973 using neutrino beams which revealed the group
structure of the electroweak SM, to the discovery of the second and
third generation (charm and tau) at SLAC 1974/1976 in
electron-positron collision which revealed the SM flavor
structure. Besides these ground-breaking discoveries, there were more
very important discoveries and achievements of $e^+ e^-$ colliders:
the first jet physics in eletron-positron collisions 1978 at the PETRA
ring at DESY which led to the discovery of the gluon, the discovery of
$B^0 - \bar{B}^0$ oscillations at the ARGUS experiment at DESY 1987
which for the first time gave proof for a top quark mass beyond 100
GeV, and the electroweak precision data from SLC and LEP at SLAC and
CERN 1989-1999 which established a (most likely) light Higgs boson
well below 200 GeV.

In the history of particle physics, there always was a fruitful
interplay between hadron and lepton machines which profited from each
other tremendously. Examples are the (guaranteed) discovery of the weak
gauge bosons $W$ and $Z$ at the CERN Sp$\overline{\text{p}}$S
proton-antiproton collider operating 0.54 TeV which was prepared by
many measurements at lepton machines, particularly by neutrino
beams. Next, there was the (guaranteed) discovery of the top quark
again in proton-antiproton collisions at the Fermilab Tevatron at
1.8/1.96 TeV, prepared by $e^+e^-$ data from DORIS/SLC/LEP at 0.01 TeV
and 0.091 TeV. They provided the mass range for the top quark which
together with the theory predictions for the QCD cross sections gave
the guarantee for a successful discovery. Then, there was the
discovery of the Higgs boson at LHC operating at 7/8 TeV (now also 13
TeV) which was prepared by the measurements of SLC and LEP I/II
operating at 0.09 TeV and 0.21 TeV. This was not as guaranteed as the
two beforementioned discoveries, but assuming the non-existence of
vastly different electroweak sectors (which would have led to
spectacular, unexpected discoveries at LHC in the first run), it was
at the same level of security than the other two.

Particle physics at the moment is at a stage where there is only a
single running high-energy collider (for the first time in at least
more than 40 years) world-wide, and that there is no
guarantee for a future discovery at any technically feasible collider
experiment in the near future. There are many preparations for a
high-energy hadron collider beyond the LHC that could energies of 60,
80 or even 100 TeV. However, also such a collider should have a
preparation by measurements at an $e^+e^-$ collider of energies in the
range of 0.35 TeV, 0.5 TeV, or 1 TeV. This would be the ILC with its
physics programme. Another issue connected to this which has not been
too much stressed up to now, are the benefits of precision QCD
measurements at the ILC, which include the strong coupling constants,
but more importantly fragmentation functions, especially for bottom
and charm. 


\section{The quest for New Physics}

The shortcomings of the SM have been described in great detail in many
publications and reports, so I will not repeat them here: the missing
dark matter particle(s), the arbitrariness of the flavor sector, the
hierarchy and fine-tuning problem, missing CP violation. As stated
already above, it is not clear whether any of these questions could be
solved at the existing or any planned collider experiments. But
clearly, there are good experiments that make it likely that there
will be the opportunity to make important discoveries even if there is
no solid guarantee as in the cases of the last section.

In the following, I will first list conditions or scenarios or cases
for possible discoveries of new physics at the ILC, compare identical
or at least similar cases from the past and try to give conditions for
possible predictions for such cases:
\begin{itemize}
\item
  {\bf New particle in kinematic reach} \\
  New particles beyond the SM are in the direct kinematic reach of the
  ILC. An example from the past of such a scenario is the charm
  discovery. It is difficult to predict, however, as one needs either
  a new symmetry principle, together with a coupling strength, or some
  indirect evidence from elsewhere like e.g. for dark matter. A future
  example at the ILC would be the discovery of an electroweakino in
  SUSY models.
\item
  {\bf New physics in (rare) decays of known particles} \\
  Existing particles will be scrutinized at the ILC, and new physics
  shows up in their rare decays. Examples from the past are anomalies
  in rare $B$ decays, future examples of this kind would be anomalies
  in (rare) Higgs decays. Again, this is difficult to predict because
  it needs tremendous technical knowledge of known physics. 
\item
  {\bf Deviations within existing interactions} \\
  An example for this would be the anomalies in the $e^+e^- \to$
  hadrons at SLC 1973 below the charm threshold, an example in the
  future would be a modification of two-fermion processes due to a
  $Z'$. Again, this is difficult to predict as it needs either a
  theoretical hint, or an experimental hint from somewhere else. 
\item
  {\bf Decipher structure of new but known interactions} \\  
  An example for this is the gluon discovery as a massless carrier of
  a confining theory, a future example would be the discovery of Higgs
  self-interactions to decipher the electroweak symmetry
  breaking. This always has guidance from existing experimental data,
  but the correct theory framework needs to be known. 
\item
  {\bf Discovery of new strong interactions} \\
  This clearly only works in the case of non-perturbative
  physics. Prime example from the past is the discovery of quark
  substructure, future examples would be discovery of (Higgs)
  compositeness. As this depends on non-perturbative physics,
  predictions are not easy or straightforward. 
\end{itemize}
The first scenario might sound trivial, but highlights another very
important for a lepton collider: namely the only guaranteed way to
discover (or exclude) any kind of weakly interacting particles up to
the scale of (half) the collider energy. This is not possible at a
hadron collider, where electroweak production processes are suppressed
sometimes even by orders of magnitude compared to strong production
processes. Furthermore, all searches at hadron colliders are subject
to model assumptions that come from trigger constraints, fiducial
phase space volume constraints to suppress tremendous backgrounds, and
also large(r) theoretical uncertainties from PDFs and missing
higher-order calculations. The scan for new weakly interacting
\begin{figure}
  \includegraphics[width=.55\textwidth]{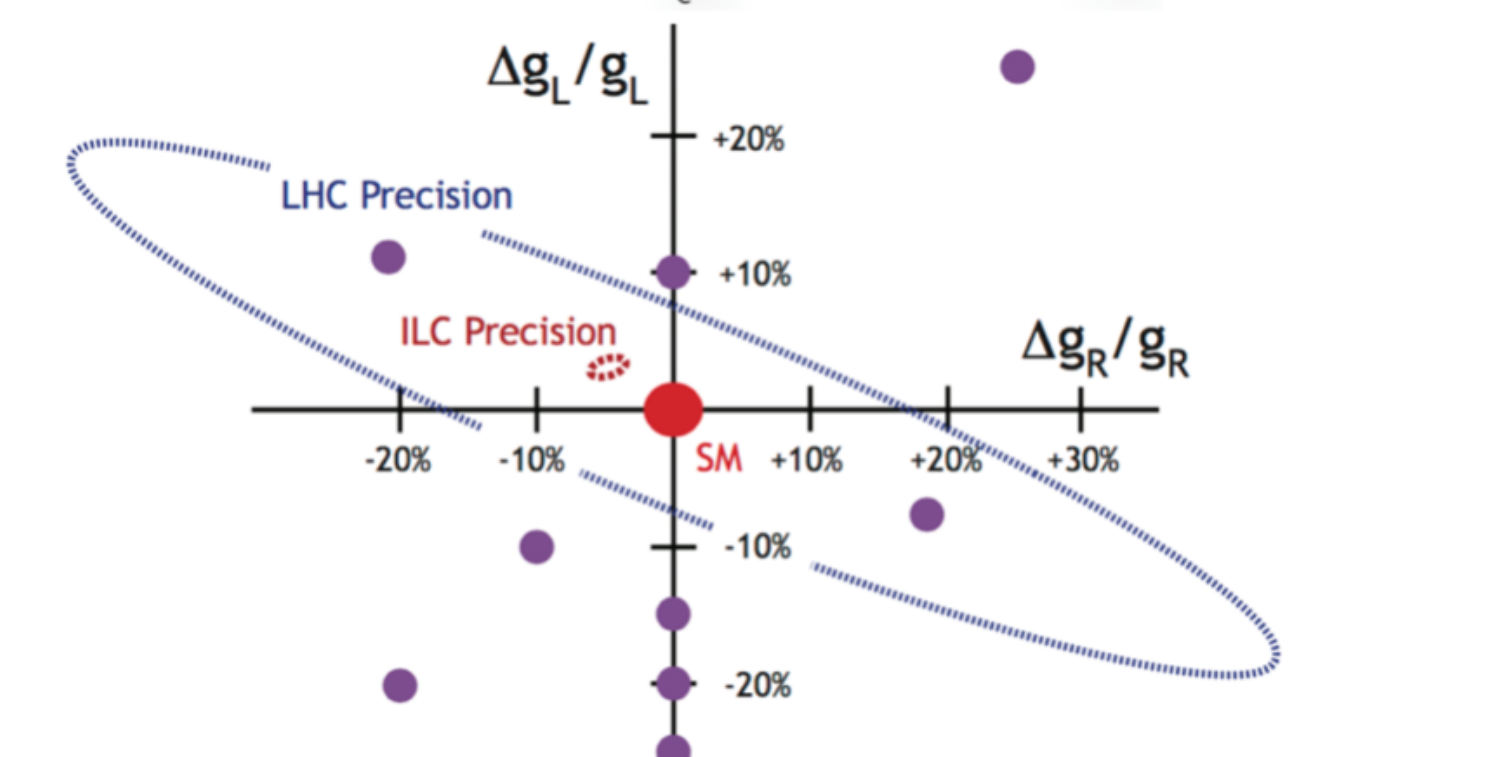}
  \includegraphics[width=.4\textwidth]{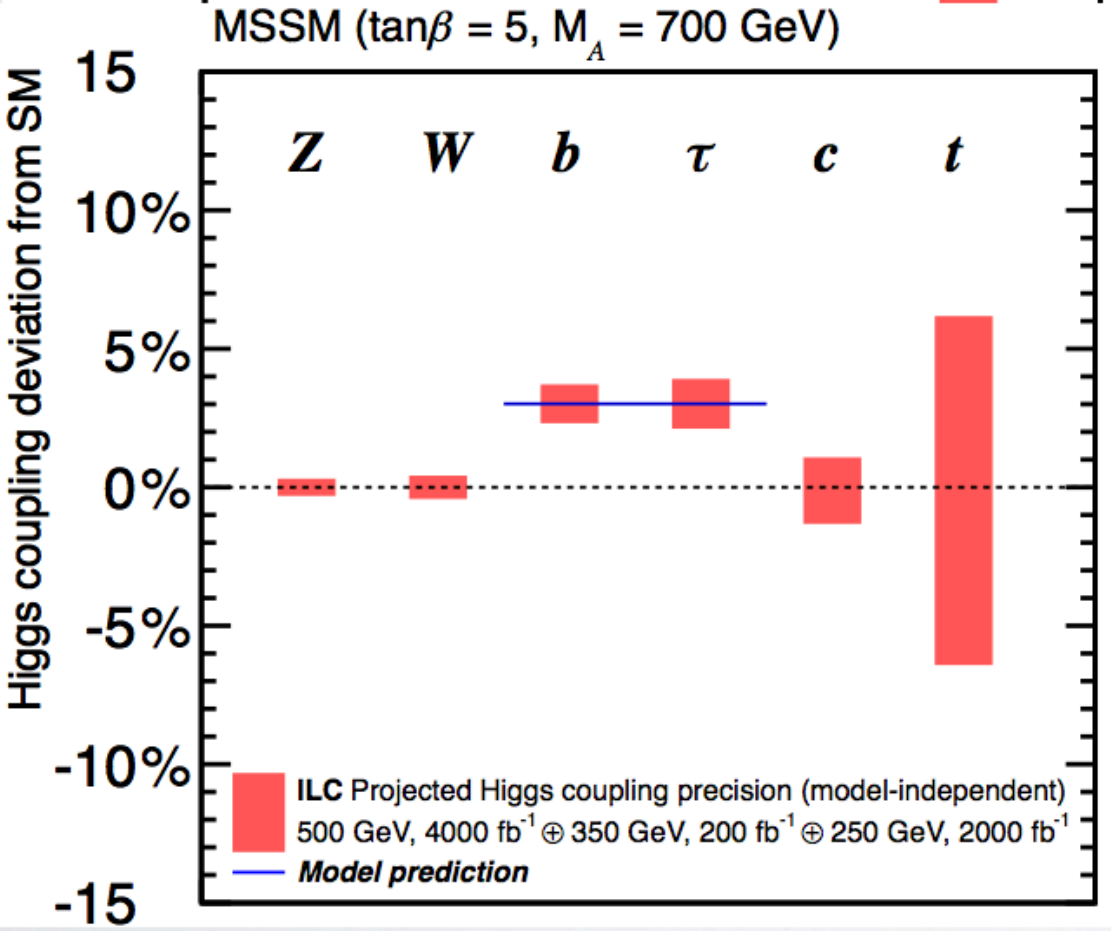}
  \caption{\label{fig:coup}
    Left plot: Collection of predictions from several BSM
    models on the left- and right-handed $ttZ$ coupling, and the
    projections with which they will be measured at LHC and
    ILC. Right plot: deviations of Higgs couplings from their SM
    values in a typical BSM scenario (composite Higgs), overlaid with
    the projected ILC precision for their measurement.}
\end{figure}
particles is the closest to a no-lose theorem for any upcoming
collider given our present knowledge on particle physics, that I am
aware of.

The two first main pillars of lepton collider physics, the top quark
and the Higgs boson, both serve as indirect tools for new physics
searches. Using an effective-field theory setup, anomalous top quark
couplings and anomalous Higgs couplings parameterize new physics in
both sectors most relevant for electroweak symmetry breaking in a way
as model-independent as it can get. Constrained by electroweak
precision data and direct measurements from Tevatron and LHC,
deviations from their corresponding SM values are expected to be in
the range of up to 10 or 20 per cent for top couplings and up to five
per cent for Higgs couplings. Fig.~\ref{fig:coup}
(from~\cite{Richard:2014upa}) shows on the 
left-hand side a collection of typical BSM models and their deviations
for the left- and right-handed top couplings to the $Z$ boson. The
ellipses mark the projections for the final LHC precision and the
expected ILC precision on these parameters. Hence, though a deviation
from the SM might have been already established at the LHC, only the
resolution power of the ILC can reveal which is the underlying theory
of Nature. In certain models, a sensitivity towards new physics scale
of the order of 50-60 TeV is reached. The right plot in
Fig.~\ref{fig:coup} denotes a typical BSM scenario, namely one of the
so-called minimal compositeness models, and its influence on the Higgs
couplings normalized to their SM value. Again, this shows that a
resolution power is needed that allows for a determination of the
Higgs coupling at the per cent level at least, as it is provided at
the ILC. 

Both the measurements of the coupling of the Higgs boson to top quarks
and to itself are among the most crucial measurements as searches for
new physics in the electroweak sector. The ILC will reach higher
precision in their measurements than the LHC even in its
high-luminosity phase. 

In principle, at the same footing are the measurements of the
properties of the electroweak gauge bosons, $W$ and $Z$. Also there
measurements at lepton colliders can trade precision frontier into
energy frontier, but this usually demands for special setups like
running at the $Z$ or $WW$ thresholds or high-energy runs at 1 TeV or
beyond~\cite{Beyer:2006hx,vv_clic}. New physics in the pure
electroweak gauge/Higgs sector is not much constrained from LHC data
at the moment, and the reach at the LHC is limited as these
measurements are complicated~\cite{vv_eff_lhc}.

Turning to direct searches for BSM at the ILC, the main driving horse
are searches for dark matter particles, in the form of mono-photon
searches similar to the mono-jet searches at the LHC. Polarization of
both electron and positron beams are inevitable tools to both enhance
the signal and suppress severe backgrounds ($W$ exchange for the
$e^+e^- \to \nu\nu\gamma$ SM background). The signal will be visible
in the structure of the high-energy part of the recoil spectrum of the
photon, cf.~e.g.~\cite{Bartels:2012ex}. A nice complementarity in
coverage between LHC and ILC have been found, in the sense that LHC
covers higher masses for dark matter particles, while ILC in general
covers far smaller couplings or higher mediator
masses~\cite{Fujii:2015jha}. In Ref.~\cite{Berggren:2013vfa}, studies
showed how regions of supersymmetry parameter space that are
inaccessible at the LHC (because they have e.g. almost degenerate LSP
and NLSP states with a mass difference of the order 1 GeV or below).
The ILC could nicely resolve such a degenerate spectrum and would
allow in such a case the reconstruction of high-scale SUSY parameters
with a precision of the order of five per cent. That the ILC can cover
the whole SUSY spectrum possibilities have been proved in a scan over
all most likely NLSP candidates ($\tilde{\mu}$ and $\tilde{\tau}$)
in~\cite{Berggren:2013vna}, where it is again obvious that the ILC can
make use of its whole phase space and detect LSP/NLSP particles very
close to the kinematic limit (as close as 2 to 10 GeV in the LSP-NLSP
mass plane). This works even for NLSP particles like sneutrinos that
are themselves invisible~\cite{Kalinowski:2008fk}.

One prime example for new weakly interacting particles that are
difficult or even impossible to detect at hadron machines like the LHC
are light scalar or pseudoscalar pseudo-Nambu-Goldstone bosons of
spontaneously broken global $U(1)$ symmetries. They decay
predominantly to the heaviest possible SM fermions, while having only
anomaly-induced couplings to diphotons or digluons (the 750 GeV
anomaly at the LHC from December 2015 has fostered interest in these
particles again). At the LHC these particles (which have similar
properties than light Higgs singlet admixtures) are only detectable if
they are heavier than approximately 200 GeV, as they cannot couple as
strongly to diphotons and -gluons as the SM Higgs boson due to light
constraints and precision data. At the ILC they can be searched for as
resonances in the $ttbb$ invariant mass spectrum as they are radiated
from top quarks, even if they are in the mass range of 10-200
GeV~\cite{Kilian:2004pp,Kilian:2006eh}.

As a last topic let me mention the search for new neutral current,
which was one of the reasonings for high-energy lepton colliders since
the 1980s. $Z'$ bosons from $U(1)$ or also non-Abelian extensions of
the electroweak gauge sector can be detected in two-fermion processes
at lepton colliders, cf.~\cite{Osland:2009dp}. This allows a
sensitivity to new physics scales reaching as far up as 100
TeV~\cite{Battaglia:2001fr}. Using the interference between $Z$ and
$Z'$, even some information on the structure of possible GUT groups at
scales of $10^{12}-10^{13}$ TeV can be gained
(cf. e.g.~\cite{Braam:2010sy}). Again, from the left- and right-handed
couplings extracted from forward-backward asymmetries and charge
asymmetries in two-fermion processes, different high-scale models can
be discriminated (cf.~e.g.~\cite{Godfrey:2005pm}).

One final remark: if something similar like the 2 TeV anomaly in $WW/WZ/ZZ$ at
the end of the 8 TeV run or the 750 GeV anomaly in diphotons will
remain at the end of run II or the high-lumi run, then the ILC is the
only option in the near future to comfirm or refute such a signal.


\section{Summary}

In this talk I tried to collect the facts in favor of a future
high-energy lepton collider (that is capable to reach at least 500
GeV) with the focus lying on new physics beyond the SM. Both the two
main SM pillars, the Higgs boson and top quark measurements serve as
indirect tools for new physics searches, but there is also a plethora
of direct search opportunities at such a machine. Most prominent
examples are dark matter searches, searches for other light weakly
coupling particles, and a scan over all weakly interacting particles.
The interplay of the ILC with the LHC, but more importantly with
future hadron machines is elucidated. Conditions, or better, scenarios
for possible BSM discoveries at the ILC have been given. Several prime 
examples for the BSM potential of the ILC have been highlighted. 

\section*{Acknowledgments}
JRR wants to thank the organizers for a fantastic conference in the
Canadian wilderness. Many thanks who contributed the content of this
discussion, among them are J.~Bagger, M.~Berggren, J.~Kanlinowski,
W.~Kilian, J.~List, J.~Mnich, M.~Peskin, F.~Richard,
G.~Wilson. Special thanks go to P.~Zerwas for guiding all of us in the
field of lepton collider physics over decades.   



\baselineskip15pt
\begin{thebibliography}{99}

\bibitem{Hagiwara:1987df} 
  K.~Hagiwara and K.~Hidaka,
  {\em Physics at TeV Energy Scale. Proceedings, Meeting, Tsukuba,
    Japan, May 28-30, 1987}, 
  KEK-87-20;
  K.~Hagiwara and S.~Komamiya,
  {\em Search For New Particles at $e^+e^−$ Colliders}, 
  Adv.\ Ser.\ Direct.\ High Energy Phys.\  {\bf 1}, 785 (1988).
\bibitem{Tracas:1987ud} 
  N.~D.~Tracas and P.~M.~Zerwas,
  {\em $e^+ e^-$ Colliders: The Window To Z's Beyond The Total
    Energy}, In *La Thuile/Geneva 1987, Proceedings, Physics at future
  accelerators, vol. 2* 214-219.    
\bibitem{Peskin:1988cz} 
  M.~E.~Peskin,
  {\em Theory of $e^+e^-$ collisions at very high energy},
  Conf.\ Proc.\ C {\bf 8708101}, 1 (1987).

\bibitem{Richard:2001qm} 
  F.~Richard, J.~R.~Schneider, D.~Trines and A.~Wagner,
  hep-ph/0106314.
\bibitem{AguilarSaavedra:2001rg} 
  J.~A.~Aguilar-Saavedra {\it et al.} [ECFA/DESY LC Physics Working Group Collaboration],
  hep-ph/0106315.
\bibitem{Behnke:2013xla} 
  T.~Behnke {\it et al.},
  arXiv:1306.6327 [physics.acc-ph].
\bibitem{Baer:2013cma} 
  H.~Baer {\it et al.},
  {\em The International Linear Collider Technical Design Report -
    Volume 2: Physics}, 
  arXiv:1306.6352 [hep-ph].
\bibitem{Baer:2013vqa} 
  H.~Baer, M.~Berggren, J.~List, M.~M.~Nojiri, M.~Perelstein,
  A.~Pierce, W.~Porod and T.~Tanabe, 
  {\em Physics Case for the ILC Project: Perspective from Beyond the
    Standard Model}, 
  arXiv:1307.5248 [hep-ph].
\bibitem{Fujii:2015jha} 
  K.~Fujii {\it et al.},
  {\em Physics Case for the International Linear Collider}, 
  arXiv:1506.05992 [hep-ex].
\bibitem{Brau:2015ppa} 
  J.~E.~Brau {\it et al.} [ILC Parameters Joint Working Group Collaboration],
  arXiv:1510.05739 [hep-ex].
\bibitem{Degrassi:2012ry} 
  G.~Degrassi, S.~Di Vita, J.~Elias-Miro, J.~R.~Espinosa, G.~F.~Giudice, G.~Isidori and A.~Strumia,
  {\em Higgs mass and vacuum stability in the Standard Model at NNLO}, 
  JHEP {\bf 1208}, 098 (2012)
  doi:10.1007/JHEP08(2012)098
  [arXiv:1205.6497 [hep-ph]].
\bibitem{Richard:2014upa} 
  F.~Richard,
  {\em Present and future constraints on top EW couplings}, 
  arXiv:1403.2893 [hep-ph].
\bibitem{Beyer:2006hx} 
  M.~Beyer, W.~Kilian, P.~Krstonosic, K.~Monig, J.~Reuter, E.~Schmidt and H.~Schroder,
  {\em Determination of New Electroweak Parameters at the ILC -
    Sensitivity to New Physics}, 
  Eur.\ Phys.\ J.\ C {\bf 48}, 353 (2006)
  doi:10.1140/epjc/s10052-006-0038-0
  [hep-ph/0604048].
\bibitem{vv_clic}
  C.~Fleper, W.~Kilian, T.~Ohl, J.~Reuter
  {\em High-Energy Vector Boson Scattering at CLIC},
  in preparation.
\bibitem{vv_eff_lhc}
  A.~Alboteanu, W.~Kilian and J.~Reuter,
  {\em Resonances and Unitarity in Weak Boson Scattering at the LHC}, 
  JHEP {\bf 0811}, 010 (2008)
  doi:10.1088/1126-6708/2008/11/010
  [arXiv:0806.4145 [hep-ph]];
  W.~Kilian, T.~Ohl, J.~Reuter and M.~Sekulla,
  Phys.\ Rev.\ D {\bf 91}, 096007 (2015)
  doi:10.1103/PhysRevD.91.096007
  [arXiv:1408.6207 [hep-ph]];
  W.~Kilian, T.~Ohl, J.~Reuter and M.~Sekulla,
  Phys.\ Rev.\ D {\bf 93}, no. 3, 036004 (2016)
  doi:10.1103/PhysRevD.93.036004
  [arXiv:1511.00022 [hep-ph]].
\bibitem{Bartels:2012ex} 
  C.~Bartels, M.~Berggren and J.~List,
  {\em Characterising WIMPs at a future $e^+e^-$ Linear Collider}, 
  Eur.\ Phys.\ J.\ C {\bf 72}, 2213 (2012)
  doi:10.1140/epjc/s10052-012-2213-9
  [arXiv:1206.6639 [hep-ex]].
\bibitem{Berggren:2013vfa} 
  M.~Berggren, F.~Brümmer, J.~List, G.~Moortgat-Pick, T.~Robens, K.~Rolbiecki and H.~Sert,
  {\em Tackling light higgsinos at the ILC}, 
  Eur.\ Phys.\ J.\ C {\bf 73}, no. 12, 2660 (2013)
  doi:10.1140/epjc/s10052-013-2660-y
  [arXiv:1307.3566 [hep-ph]].
\bibitem{Berggren:2013vna} 
  M.~Berggren,
  {\em Simplified SUSY at the ILC}, 
  arXiv:1308.1461 [hep-ph].
\bibitem{Kalinowski:2008fk} 
  J.~Kalinowski, W.~Kilian, J.~Reuter, T.~Robens and K.~Rolbiecki,
  {\em Pinning down the Invisible Sneutrino}, 
  JHEP {\bf 0810}, 090 (2008)
  doi:10.1088/1126-6708/2008/10/090
  [arXiv:0809.3997 [hep-ph]].
\bibitem{Kilian:2004pp} 
  W.~Kilian, D.~Rainwater and J.~Reuter,
  {\em Pseudo-axions in little Higgs models}, 
  Phys.\ Rev.\ D {\bf 71}, 015008 (2005)
  doi:10.1103/PhysRevD.71.015008
  [hep-ph/0411213].
\bibitem{Kilian:2006eh} 
  W.~Kilian, D.~Rainwater and J.~Reuter,
  {\em Distinguishing little-Higgs product and simple group models at
    the LHC and ILC}, 
  Phys.\ Rev.\ D {\bf 74}, 095003 (2006)
  [Phys.\ Rev.\ D {\bf 74}, 099905 (2006)]
  doi:10.1103/PhysRevD.74.095003, 10.1103/PhysRevD.74.099905
  [hep-ph/0609119].
\bibitem{Osland:2009dp} 
  P.~Osland, A.~A.~Pankov and A.~V.~Tsytrinov,
  {\em Identification of extra neutral gauge bosons at the
    International Linear Collider}, 
  Eur.\ Phys.\ J.\ C {\bf 67}, 191 (2010)
  doi:10.1140/epjc/s10052-010-1272-z
  [arXiv:0912.2806 [hep-ph]].
\bibitem{Battaglia:2001fr} 
  M.~Battaglia, S.~De Curtis, D.~Dominici and S.~Riemann,
  eConf C {\bf 010630}, E3020 (2001)
  [hep-ph/0112270].
\bibitem{Braam:2010sy} 
  F.~Braam, A.~Knochel and J.~Reuter,
  JHEP {\bf 1006}, 013 (2010)
  doi:10.1007/JHEP06(2010)013
  [arXiv:1001.4074 [hep-ph]].
\bibitem{Godfrey:2005pm} 
  S.~Godfrey, P.~Kalyniak and A.~Tomkins,
  {\em Distinguishing between models with extra gauge bosons at the
    ILC}, 
  hep-ph/0511335.
\end{thebibliography}
\end{document}